\documentstyle [epsfig]{mn}

\newif\ifAMStwofonts

\ifoldfss
  \ifCUPmtlplainloaded \else
    \NewTextAlphabet{textbfit} {cmbxti10} {}
    \NewTextAlphabet{textbfss} {cmssbx10} {}
    \NewMathAlphabet{mathbfit} {cmbxti10} {} 
    \NewMathAlphabet{mathbfss} {cmssbx10} {} 
  \fi
  \ifAMStwofonts
    \ifCUPmtlplainloaded \else
      \NewSymbolFont{upmath} {eurm10}
      \NewSymbolFont{AMSa} {msam10}
      \NewMathSymbol{\upi}     {0}{upmath}{19}
      \NewMathSymbol{\umu}     {0}{upmath}{16}
      \NewMathSymbol{\upartial}{0}{upmath}{40}
      \NewMathSymbol{\leqslant}{3}{AMSa}{36}
      \NewMathSymbol{\geqslant}{3}{AMSa}{3E}

      \let\leq=\leqslant \let\le=\leqslant
      \let\geq=\geqslant 
    \fi
  \fi
\fi 

\ifnfssone
  \newmathalphabet{\mathit}
  \addtoversion{normal}{\mathit}{cmr}{m}{it}
  \addtoversion{bold}{\mathit}{cmr}{bx}{it}
  \newmathalphabet{\mathbfit} 
  \addtoversion{normal}{\mathbfit}{cmr}{bx}{it}
  \addtoversion{bold}{\mathbfit}{cmr}{bx}{it}
  \newmathalphabet{\mathbfss} 
  \addtoversion{normal}{\mathbfss}{cmss}{bx}{n}
  \addtoversion{bold}{\mathbfss}{cmss}{bx}{n}
  \ifAMStwofonts
    \ifCUPmtlplainloaded \else
      \UseAMStwoboldmath
      \makeatletter
      \new@mathgroup\upmath@group
      \define@mathgroup\mv@normal\upmath@group{eur}{m}{n}
      \define@mathgroup\mv@bold\upmath@group{eur}{b}{n}
      \edef\UPM{\hexnumber\upmath@group}
      \new@mathgroup\amsa@group
      \define@mathgroup\mv@normal\amsa@group{msa}{m}{n}
      \define@mathgroup\mv@bold\amsa@group{msa}{m}{n}
      \edef\AMSa{\hexnumber\amsa@group}
      \makeatother
      \mathchardef\upi="0\UPM19
      \mathchardef\umu="0\UPM16
      \mathchardef\upartial="0\UPM40
      \mathchardef\leqslant="3\AMSa36
      \mathchardef\geqslant="3\AMSa3E

      \let\leq=\leqslant \let\le=\leqslant
      \let\geq=\geqslant 
    \fi
  \fi
\fi 

\ifnfsstwo
  \DeclareMathAlphabet{\mathbfit}{OT1}{cmr}{bx}{it}
  \SetMathAlphabet\mathbfit{bold}{OT1}{cmr}{bx}{it}
  \DeclareMathAlphabet{\mathbfss}{OT1}{cmss}{bx}{n}
  \SetMathAlphabet\mathbfss{bold}{OT1}{cmss}{bx}{n}
  \ifAMStwofonts
    \ifCUPmtlplainloaded \else
      \DeclareSymbolFont{UPM}{U}{eur}{m}{n}
      \SetSymbolFont{UPM}{bold}{U}{eur}{b}{n}
      \DeclareSymbolFont{AMSa}{U}{msa}{m}{n}
      \DeclareMathSymbol{\upi}{0}{UPM}{"19}
      \DeclareMathSymbol{\umu}{0}{UPM}{"16}
      \DeclareMathSymbol{\upartial}{0}{UPM}{"40}
      \DeclareMathSymbol{\leqslant}{3}{AMSa}{"36}
      \DeclareMathSymbol{\geqslant}{3}{AMSa}{"3E}

      \let\leq=\leqslant \let\le=\leqslant
      \let\geq=\geqslant 
    \fi
  \fi
\fi 

\ifCUPmtlplainloaded \else
  \ifAMStwofonts \else 
    \def\upi{\pi}
    \def\umu{\mu}
    \def\upartial{\partial}
  \fi
\fi

\title[The dwarf LSB galaxy population of the Virgo Cluster I.]
{The dwarf LSB galaxy population of the Virgo Cluster I. \newline
The faint-end slope of the Luminosity Function.}
\author[S. Sabatini et al.]
       {S. Sabatini$^1$, J. Davies$^1$, R. Scaramella$^2$, R. Smith$^1$, M. Baes$^{1,3}$ \thanks{Postodoctoral Fellow of the Fund for Scientific Research, Flanders Belgium (FWO-Vlaanderen)}, 
	S. M. Linder$^1$, 
 \newauthor    S. Roberts$^1$,   V. Testa$^2$  \\
        $^1$Cardiff University, Cardiff, UK \\ $^2$INAF-OAR, Rome, Italy \\
	$^3$Sterrenkundig Observatorium, Gent, Belgium }
\date{Accepted 1988 December 15.
      Received 1988 December 14;
      in original form 2002 May 30}

\pagerange{\pageref{firstpage}--\pageref{lastpage}}
\pubyear{2002}

\begin{document}

\maketitle

\label{firstpage}


\begin{abstract}
The widely varying dwarf galaxy counts in different environments
provide a strong challenge to standard hierarchical clustering
models. The luminosity function is not universal, but seems to be
strongly dependent upon environment. In this paper we describe an
automated procedure for detecting and measuring very low surface
brightness (LSB) features in deep CCD data. We apply this procedure to
large area CCD survey fields of the Virgo cluster. We show that
there are many more faint ($-10 \geq M_{B} \geq -14$) low surface
brightness galaxies than what would be predicted from
extrapolation of the Virgo cluster catalogue luminosity function.
Over our limited range of measurement the faint end slope of the
luminosity function becomes $\alpha=- 1.6$. The luminosity
function is flatter in the inner regions of the cluster than it is
in the outer regions. Although these galaxies contribute a small
fraction of the total stellar light of the cluster, they may
contribute significantly to the mass in galaxies if they have
large mass-to-light ratios similar to those recently measured for
Local Group dwarf galaxies.
\end{abstract}

\begin{keywords}
dwarf galaxies -- virgo cluster -- luminosity function.
\end{keywords}


\section{Introduction}

The Virgo cluster offers the best opportunity to study in detail large numbers
of galaxies (also dwarf galaxies) over a small region of sky. 
It is the nearest (d$\sim17$ Mpc, Tikhonov et al., 2000)
 cluster with several hundreds of galaxies ($\sim 1277$ members, Binggeli et al, 1984)
 and the largest dominant structure of the Local Supercluster. It is an irregularly
shaped cluster, with a high abundance of spiral galaxies among the
bright cluster members and a large population (80\% of the total
known galaxy number) of dwarf galaxies (Binggeli et al., 1985). Its size is
approximately 10 degrees across the sky, which corresponds to
$\sim 3$ Mpc. The crossing time for this cluster is $\approx 0.1
H_0^{-1}$ (Trentham et al., 2002) so that cluster galaxies
have had plenty of opportunities to interact with each other. The
huge mass of the cluster ($M=1.2 \times 10^{15} M_{\odot}$,
Fouque, 2001) accelerates the member galaxies to very high
 peculiar velocities, so that it exhibits the highest blue-shift
measured for any galaxy (IC3258, approaching us at 1600 km/sec).
As suggested from its irregular structure, the cluster is made up
of at least three subclusters dominated by the bright ellipticals
M87, M86, M49 and thus it is probably not in a state of
equilibrium: it's a very complex unrelaxed system where the
central part is dynamically old, while the large surrounding
region is not virialized, but still infalling (Bohringer, 1995).

The first observations of  the cluster date back to Mechain 
and Messier (late $18^th$ century), who noticed a large concentration of nebulae
in the northern wing of the Virgo constellation. The
identification of the cluster as a self-gravitating system
consisting of hundreds of galaxies
 followed soon after Hubble's 1923 discovery of Cepheids in M31. The first systematic
investigation of the cluster was carried out by Shapley and Ames (1932). Since then
the Virgo cluster has been of primary importance for extragalactic astronomy.
It  has been subject to many studies at optical, radio, IR, and X-ray wavelengths
aimed at addressing its galaxy population, evolution and gas content (for the most recent
studies see Gavazzi et al, 2002 (optical \& multi$\lambda$); Van Driel et al., 2000 (radio)
Tuffs et al., 2002 (IR); Shibata et al, 2001 (X-ray)).

The most complete optical survey of the Virgo cluster
was carried out by Binggeli et al. in 1984, using
photographic plates and a visual detection method. Whilst the advantage
of using photographic plates is the  wide area of sky surveyed
($\approx 6^{\circ}$ around the center of the cluster in this
case), the major disadvantage is the low sensitivity: the survey is
incomplete for objects with $M_{B}\le-14$, missing the numerous
galaxies that dominate the numbers in the Local
Group, for example.  Binggeli et al. (1984) carried out an extensive
systematic study of the cluster finally producing a catalogue of
2096 galaxies (the Virgo Cluster Catalogue, VCC). The papers produced by them
contain photometry and morphology for the galaxies, analyze morphology
and dynamics and study specific and general luminosity functions
for the cluster (Binggeli et al.,1984, 1985, 1987; Sandage
et al., 1984, 1985, 1985a).

Our main concern in this paper is with the numbers of faint Virgo
dwarf galaxies that the subjective, visual detection method used by Binggeli et al. and
others might have missed in less deep surveys. The observed galaxy counts are quantified by
a determination of the galaxy Luminosity Function, LF,
(described by a Schechter function) and particularly
from the value of its faint end slope, $\alpha$. After correction for
incompleteness, Binggeli et al. found a value for the faint
end slope of $\alpha \sim -1.35$.
Following on from the Binggeli et al. survey, Impey et al. 
(1988) used a photographic amplification technique to reach
lower surface brightnesses. They found numerous additional LSB
dwarf galaxies not included in the VCC and derived a value of
$\alpha \sim -1.7$. Using the same technique they also obtained a
steepening in the faint end slope of the Fornax cluster from -1.3 to -1.55 
(Bothun et al., 1991). \newline
A more recent study of the Virgo cluster by
Phillipps et al., 1998 (again using photographic plates)
produced a steeper slope again ($\alpha \sim -2.0$).
 Phillipps et al. used a statistical method to obtain the 
luminosity function, subtracting faint
galaxy number counts of fields outside of the cluster from those
containing the cluster. This technique is very delicate because
two large numbers are being subtracted away from each other to
leave a small residual.
There are also large variations in the background counts 
(Valotto et al. 2001). These rather steep values for the faint 
end slope of the luminosity function of Virgo are consistent with some 
values obtained for other clusters. For example Bernstein et al. 
(1995) found a value of $\alpha \sim -1.4$ for the more distant Coma cluster 
while Kambas et al. (2000) found a very steep slope of $\alpha \sim -2.0$ for the 
nearby Fornax cluster.

 These values differ, however, from  recently determined values of the faint end slope of the field
galaxy population derived from the extensive 2dF and Sloan
surveys, where $\alpha \sim -1.2$ (Cross et al., 2001, Blanton et
al., 2001). The field galaxy luminosity function also corresponds
very well with that obtained for the Local Group, $\alpha \sim
-1.1$ (Mateo, 1998).

 A major concern with the measurement of the dwarf galaxy content in 
different environments, is the wide range of data and detection methods used 
by different groups and how this affects the derived value of the faint end
 slope. We have recently looked at fields in different
 environments (Sabatini et al, JENAM 2002, Roberts et al., in prep.) 
using data in exactly the same conditions to that described below for Virgo.
Trentham et al. (2002) found that the luminosity function is
strongly dependent upon environment and, as a preliminary result, we
find very few dwarf galaxies in the field and around isolated
galaxies, if compared with numbers in Virgo.

It is clear that an environmental dependency of the luminosity
function places new and challenging constraints on current
theories of galaxy formation. For example standard Cold Dark
Matter (CDM) theories predict steep faint end slopes in all
environments (Bullock et al., 2000). Additions to the theory, such as supernovae driven
winds (Dekel and Silk, 1986) or the influence of a reheated
intergalactic medium (Efstathiou, 1992) have been used to suppress
dwarf galaxy formation globally. These arguments cannot be used to
explain the large differences in the dwarf galaxy population
content in some environments. Recently Tully et al. (2002) have
suggested that
 dwarf galaxy numbers in different environments depend on whether the structure
 in which they reside is formed
before or after reionization ('squelching'). They compared the
Virgo cluster, which is dwarf rich, to the dwarf poor Ursa Major
cluster. From their simulations they suggest that Virgo was
assembled with its dwarf galaxy population before reionization and
Ursa Major after, so that dwarf galaxy formation was suppressed.
There is though, as they say, only 'qualitative' agreement between
the model and the observations. Moore et al. (1999) offer more of a nurture rather 
than a nature explanation to the problem. Their galaxy
'harassment' model provides a mechanism for the formation
 of dwarf galaxies in the cluster environment through disruption of discs due to
tidal interactions with the cluster and individual galaxies. The nature of the dwarf galaxy population
 is then very dependent on the dynamical properties of the cluster (crossing time, galaxy number density).
 The Virgo cluster has a very short crossing time (0.1 $H^{-1}_{o}$) compared to Ursa Major (0.5 $H^{-1}_{o}$)
(Tully et al. 2002). Thus the Virgo cluster galaxies have had more
opportunities to become 'harassed'. In a study of the dynamical
properties of Virgo cluster galaxies Conselice et al. (2001)
describe evidence supporting the idea that the dwarf galaxies are
dynamically distinct from the older cluster galaxy population 
and have thus been derived from an infalling population as envisaged by the Moore 
et al. harassment model.

An accurate derivation of the dwarf galaxy population as a
function of environment is the only way to distinguish between
these models. We need a detailed comparison of the luminosity
function derived consistently for galaxies in different
environments with those predicted by CDM, squelching and
harassment.
This is the first of a series of papers in which we will study in more detail 
the dwarf galaxy population identified in Virgo and then 
we will derive the environmentally dependent luminosity
function of galaxies for comparison with the models.

Dwarf galaxies are extremely difficult to detect because they generally
have both very low luminosity and surface brightness. To
detect objects like this we need an optimum filter to enhance the
signal and a set of selection criteria that preferentially selects
cluster members rather than background galaxies. We also require
an automated and repeatable procedure that can be applied
consistently to data taken from a large area survey.

The paper is organized as follows: in section 2 we present the
Isaac Newton Telescope (INT) Wide Field Camera (WFC) survey data
(http://www.ast.cam.ac.uk/~wfcsur/index.php). In section 3 we
discuss the methods used to determine cluster membership and to
assess possible background contamination. In section 4 we describe
the detection algorithm (see also Sabatini et al., 1999;
Scaramella et al., in prep.).
 The luminosity function and first results
are presented in section 5.


\section[]{Data}

\begin{figure}
\centerline{\epsfig{file=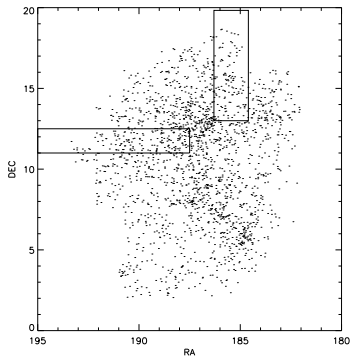,height=5
cm}}
{\caption{\footnotesize Plot of the Virgo Cluster members as
from the VCC. Overlayed are the areas covered by the INT WFCS.}}
\label{fig:Virgo}
\end{figure}

The data we use are part of the INT WFC survey, that is a multi-colour CCD based wide field
survey covering an area of $\sim 200$ deg$^{2}$. The camera is mounted at the prime
focus of the 2.5m INT on La Palma, Canary Island, it consists of 4 thinned
EEV 2kx4k CCDs with pixel size 0.33'' and has a field size of
34.2'x34.2' (neglecting the 1' inter-chip spacing). The data for
the Virgo survey were acquired during observing runs in Spring
1999 to 2002 and consist of two perpendicular strips of B and I
band CCD images extending from the center of the cluster (identified as M87) outward
for 7 and 5 degrees respectively (see fig \ref{fig:Virgo}). The
total area covered is $\sim 25$ deg$^{2}$ and the average sky
noise corresponds to $\sim 26$ B mag/ sq arcsec. With the use of the techniques
we describe in sec. \ref{sec:algo}, these deep data allow us to
study the population of dwarf galaxies down to very faint limiting
central surface brightnesses ($\sim 26 B$ mag/sq arcsec) and
absolute magnitudes ($M_B \sim -10$) (see below). The results
presented in this paper make use of the East-West strip, that
includes M87.

The data were preprocessed and fully reduced using the Wide Field Survey pipeline. This includes de-biassing,
bad pixel replacement, non-linearity correction, flatfielding, defringing (for I,Z bands)
and gain correction.  The photometric
calibration makes use of several (5-10 per night) standard stars and the zero-points
are accurate to 1-2$\%$.  For the details see http://ast.cam.ac.uk/~wfcsur/pipeline.html.
World Coordinate System (WCS) information is embedded in the reduced images
FITS header and the astrometric calibration errors are $\leq$ 1 arcsec.
The median seeing was 1.9 arcsec.


\section{Membership determination and background subtraction.}

Membership determination is one of the crucial and controversial problems 
in the study of galaxy populations in nearby clusters (Valotto et al., 2001). 
The properties of the background galaxies need to be studied
in detail in order to minimize contamination in the sample. With
the aim of deriving selection criteria which will enable us to
separate cluster from background galaxies, we have carried out
numerical simulations of the galaxy population in order to
identify distinguishing properties of member and non-member
galaxies. \newline The final goal is to find selection criteria that
minimize the background contamination in the sample and at the
same time maximize the number of Virgo cluster members selected. We have
then tested the validity of the derived selection criteria in two ways
(see ahead):
\begin{enumerate}
\item by considering the fall in surface density of our detections
with increasing distance from the cluster centre. \item by
measuring the number of background galaxies detected, using our
selection criteria, on fields outside the cluster.
\end{enumerate}


\subsection{Numerical Simulations of a cone of universe} \label{sec:numsim}
The effects of redshift on the properties of a galaxy are mainly
the following:

\begin{enumerate}
\item surface brightness dimming due to distance (with a
dependency of $(1+z)^4$ + K-corrections ); \item change in
apparent size with distance.
\end{enumerate}

As a result, in principle, an intrinsically bright distant galaxy
may appear as a faint nearby low surface brightness one and could
be included in the cluster members catalogue. For example a
$L_{*}$ galaxy ($M_{B}=-20$, $\mu_{0}$=21.7 $B\mu$) at z=0.2 has
an apparent total magnitude of 19.5 and a $\mu_{0}$ of 23 and
could be  mistaken for a cluster low luminosity LSB galaxy. Its
scale size however would be small (of order 2 arc sec) while a
cluster dwarf with the same apparent magnitude would have a
surface brightness of 24.3 $B\mu$ (because of the surface
brightness magnitude relation, see below) and hence a scale size
of order 4 arc sec. Thus in principle cluster galaxies can be
distinguished by their larger sizes and fainter surface
brightnesses at a given magnitude.

In order to quantify the trend of the apparent parameters of a
galaxy with redshift we have carried out some numerical
simulations: we adapted a code (Esslinger-Morshidi Z., PhD Thesis,
1997) that populates a cone of universe and detects the galaxies
after applying given selection criteria. The aim is to obtain an
estimate of the completeness and contamination \footnote{In what
follows by completeness we mean the ratio of the number of Virgo
members selected according to the selection criteria to total
Virgo members in the simulation. In the same way, the
contamination is given by the ratio of the number of background
galaxies selected to total number of galaxies (Virgo members +
background)} of our sample for different selection criteria. We
then use the most efficient selection criteria that preferentially
select Virgo cluster galaxies.
The code is composed of two parts: \\
1) input galaxy creation;\\
2) detection of the galaxies.\\
The cone of universe is randomly uniformly \footnote{Density fluctuations 
are not considered here, as the main goal of the simulation is to compare 
general apparent properties (morphology and photometry) of the background 
population with the Virgo Cluster galaxies.} populated in space
using a given cosmology, a Luminosity Function and a Surface
Brightness Distribution for the galaxies. The output of this first
part of the code is a
 catalogue of objects with given proper distance, apparent magnitude, central
 surface brightness and scale length (all objects are assumed to have
exponential profiles).
In the second part of the code, the given selection criteria are applied
 to the catalogue and the final output is a list of the objects
that satisfy the selection criteria along with their photometric properties.

In our simulations we chose $H_0=75$ km s$^{-1}$ Mpc$^{-1}$, a
flat universe with $\Omega_{M}=0.3$ and $\Omega_{\Lambda}=0.7$
(Lahav {\it et al.}, 2002), and we analyzed a cone from redshift
0.001 to 1.5 and size $10\times 10$ sq degrees \footnote{The large
aperture is necessary because otherwise the local universe is
underrepresented, due to the very small volume. Moreover this is
also roughly the size of the cluster}.

The luminosity function we adopted for the field is the one obtained by the 2dF Galaxy
Redshift Survey
(Madgwick et al., 2001) with the following parameters:

\begin{itemize}
\item $M_{B}*=-19.79$ \item $\alpha=-1.19$ \item $\phi_*= 0.00157$
h$^3$ Mpc$^{-3}$
\end{itemize}
and the range of input magnitudes for the galaxies is: -23 to -10.

The surface brightness distribution was chosen to follow the relationship
between Surface Brightness and Total Magnitude (for recent confirmation
see Blanton et al., 2001). Here we use the one given in
Driver, 1999:

\begin{equation}
\mu_{e} \sim [(0.67 \pm 0.08) M_{B}-(33 \pm 6)]
\end{equation}
where $\mu_{e} $ is the average surface brightness within the
effective radius and for an exponential profile:
\begin{equation}
\mu_{e} \sim \mu_{0}+1.15
\end{equation}

Using the same code with the same area (but keeping a fixed redshift of z=0.0040 $\pm$ 0.0003,
 corresponding to  the distance of the Virgo cluster, Russell et al. 2000) we generate an
artificial cluster. The Luminosity function used to describe it is
normalized to the one in Sandage et al. '85 in their survey
of the Virgo cluster:
\begin{itemize}
\item $M_*=-21.4$ B mag
\item $\phi_*= 2.36$ Mpc$^{-3}$
\end{itemize}

We leave the slope $\alpha$ as a free parameter, with the purpose
of running simulations with different values
($\alpha=-1.0,-1.2,-1.4,-1.6,-1.8,-2.0$). The surface brightness
distribution chosen to describe the Virgo cluster is the Surface
Brightness Magnitude relationship as derived by Impey et al.
(1988) for Virgo:

\begin{equation}
\mu_{0} = 0.67M+32 \pm2
\end{equation}

\begin{figure}
\centerline{\epsfig{file=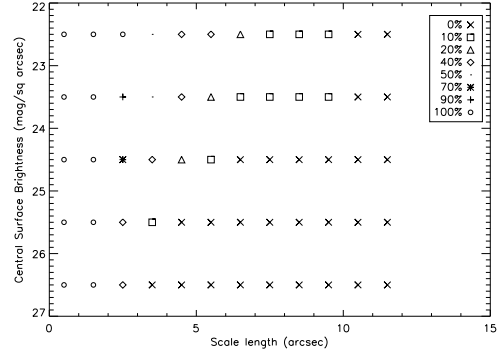,height=5
cm}} \caption{\footnotesize{ Plot of background contamination
as a function of $\mu_{0}$ (mag/sq arcsec) and
h (arcsec). The numbers refer to the simulation where
the LF has a faint end slope of -1.4 for the
Virgo Cluster.}} \label{fig:contamination}
\end{figure}

\begin{figure}
\centerline{\epsfig{file=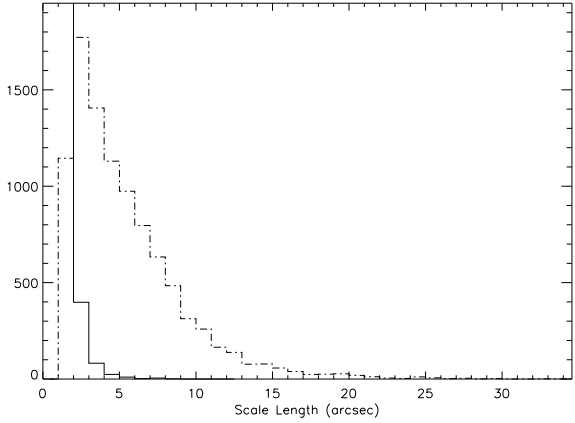,height=6
cm}} \caption{\footnotesize{ Histogram of Scale Length for both
simulations: the dashed dotted line refers to Virgo members and
the filled line to background galaxies. }} \label{fig:alfa_histo}
\end{figure}

The output catalogues produced by the code allow us to study the
behaviour of apparent central surface brightness, scale length and
total magnitude for simulations of the two different environments.
We can then estimate what is the percentage of Virgo members and
background galaxies for each bin of scale length $h$ and
central surface brightness $\mu_{0}$.  In fig.
\ref{fig:contamination} we show how the percentage contamination by
background galaxies depends on the scale length and central
surface brightness of the galaxy, for the simulation with $\alpha=-1.4$
for the Virgo LF \footnote{We show results form this simulation
because this is the flattest LF faint end slope we
would expect if we don't find any additional galaxy to the VCC ones}. 
Fig. \ref{fig:alfa_histo} shows the histogram of scale lengths for all
simulated galaxies. The dashed dotted line
refers to cluster members and the filled line to background
galaxies. Both figures clearly show that the best discriminating  
property to distinguish cluster galaxies from background ones is
the scale length:  as a general trend the background galaxies
appear to have mainly smaller sizes and this makes it possible to
separate them from Virgo members (as discussed earlier).
Selecting galaxies with scale lengths $h \geq 3 $ arcsec 
ensures that we maximize the detection of
cluster galaxies compared to background galaxies (from fig. \ref{fig:contamination}
contamination within this selection criteria is kept under 50 \%).
Being primarily interested in the low surface brightness 
population of the cluster we then restrict the selection 
to galaxies with $\mu_{0} \geq 23 $.

Simulations with LF faint end slopes for the cluster steeper than 
that of the background give a better discrimination 
between cluster and background galaxies as the cluster slope increases. 
This is illustrated in tables
\ref{fig:tab2},\ref{fig:tab3},\ref{fig:tab4}. \newline

\begin{center}
\begin{table} 
\centerline{\bf Slope of Virgo LF = -1.0}
\centering
\begin{tabular}{|c|c|c|}

\hline \hline

{\bf Scale Length cut-off} & {\bf Completeness} & {\bf Contamination} \\

\hline \hline

2 & 93\% & 62\% \\ \hline
3 & 83\% & 31\% \\ \hline
4 & 72\% & 15\% \\ \hline

\end{tabular}
\caption{ Completeness and contaminations for different choices of
the minimum scale length selected.}

\label{fig:tab2}
\centerline{\bf Slope of Virgo LF = -1.4}
\centering
\begin{tabular}{|c|c|c|}

\hline \hline

{\bf Scale Length cut-off} & {\bf Completeness} & {\bf Contamination} \\

\hline \hline

2 & 88\% & 6\% \\ \hline
3 & 70\% & 2\% \\ \hline
4 & 55\% & 0.8\% \\ \hline
\end{tabular}

\caption{ Completeness and contaminations for different choices of
the minimum scale length selected.} \label{fig:tab3}

\centerline{\bf Slope of Virgo LF = -2.0}
\centering
\begin{tabular}{|c|c|c|}

\hline \hline

{\bf Scale Length cut-off} & {\bf Completeness} & {\bf Contamination} \\

\hline \hline

2 & 82\% & 0.02\% \\ \hline
3 & 60\% & 0.00\% \\ \hline
4 & 44\% & 0.00\% \\ \hline

\end{tabular}

\caption{ Completeness and contaminations for different choices of
the minimum scale length selected.} \label{fig:tab4}

\end{table}
\end{center}

In summary our simulations have shown that in order to maximize
the ratio of Virgo members to background galaxies we should use a
selection criteria of $\mu_{o} \geq 23$ and a scale length $h$ $\geq$ 3 arcsec. 
On a practical note we found that typically the seeing on our frames was a poor 2 arc sec.
Convolving a 3 arc sec scale size galaxy with this seeing leads to
a measured scale size closer to 4 arc sec. Thus in practice we
selected objects with measured scale sizes greater than 3 arc sec
which, because of the template sizes used in the method described below, means a
minimum scale size of 4 arc sec.

\subsection {Offset fields}\label{sec:offset}

As a further check to validate our selection criteria, we have
applied our algorithm (see sec. \ref{sec:algo}) to another set of
INT WFC data using the same detector, exposure time and filter and
covering a region of sky at about the same Galactic latitude. The
data we used are part of the Millennium Galaxy Survey that is a 36
arcmin wide strip going from $9^{h}58^{m}28^{s}$ to
$14^{h}46^{m}45^{s}$, J2000 (Liske et al., 2002). From these
data we chose a number of random fields to compare our predicted
number of background detections with that of the model.  The
number of galaxies detected by our algorithm is 4 galaxies per sq
deg. This number is in agreement both with what our numerical
simulations predicted ($\sim6$ gal per sq deg) and from
 what we measure for the background counts to be in the Virgo fields. It is much less than the actual number
of galaxies ( $\approx 20$ gal/ sq deg) found in the Virgo cluster fields. All these points are discussed
 further below.


\section{The detection algorithm}\label{sec:algo}

\begin{figure}
\centerline{\mbox{\epsfig{file=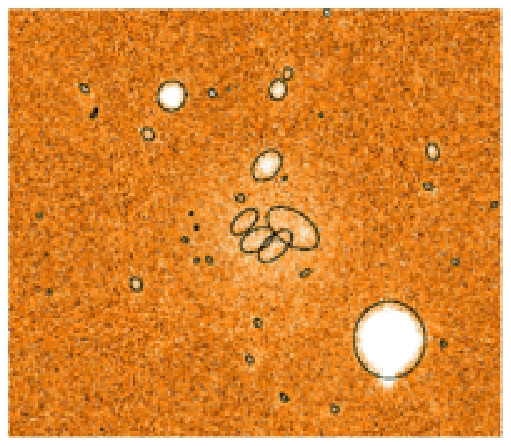,
            height=6cm}}}
\end{figure}
\begin{figure}
\centerline{\mbox{\epsfig{file=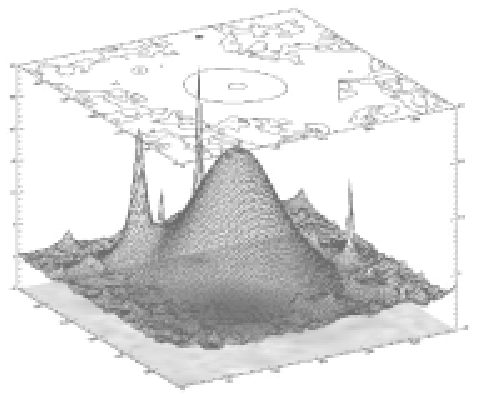,
height=6cm}}} \caption{\footnotesize { At the {\bf top} we show
the output of SExtractor for an image with a LSB galaxy from our catalogue in the centre:
as can be seen, the program classifies galaxy is identified as a group of small objects rather
than as an object as a whole. The detection in this case is difficoult because of the poor pixel-to-pixel
signal to noise ratio for this object. As a comparison at the {\bf bottom} we show the 3D output image
from our algorithm for the same field. In this image
the galaxy appears to be far above the noise level and thus easily
detectable as a single object. }} \label{fig:finding}
\end{figure}

Looking for objects with a surface flux close to the sky noise
requires the use of image enhancement
 techniques in order to optimize their detection.
Standard detection algorithms for connected pixels in fact often
fail on these kind of objects because of their poor pixel-to-pixel
signal to noise ratio (S/N). The algorithm we have developed uses
frequency domain techniques and mainly consists of convolutions of
the image with matched filters: the advantage in this case is the
use of the total flux of the galaxy to detect it, instead of the
very low S/N pixels at its edge. As an example of this, in fig.
\ref{fig:finding} we show the finding chart produced by SExtractor
(Bertin \& Arnouts, 1996) for one of our fields: a LSB galaxy
from our catalogue ($\mu_{0}=25.3$ mag/sq arcsec, $h$ =9 arcsec) is at the
centre of the image. As a comparison we also show the output of
our detection algorithm on the same image: where SExtractor finds
many distinct objects at the position of the LSB galaxies, our
procedure clearly distinguishes it as one. \\

The algorithm we used is a modified version of the method
described in Sabatini et al. (1999) \footnote{Data management and 
processing is performed using the application
package language IDL (Copyright Research System, Inc.)}
 (for more details see
Scaramella et al., in preparation) and it is composed of the
following steps:
\begin{enumerate}
\item background fluctuation flattening \item removal of standard
astronomical objects (like stars, bright galaxies, cosmic rays
etc.) \item image convolution with matched filters that are
optimized to enhance faint fuzzy structures \item candidate
classification.
\end{enumerate}

LSB and dwarf galaxy candidate identification is performed on the
final convolved image by means of selecting all peaks that are
significantly (see sec. \ref{sec:multifilt}) above the residual
noise fluctuations. The whole procedure (that includes masking,
filtering, detection and measurement) is automated and its
efficiency has been tested using
artificial galaxies added to real images (see sec. \ref{sec:artgal}).

Techniques involving convolution with matched filters have been used before 
(see for example Armandroff et al. 1998 or Flint et al. 2001) but the 
innovations of our method are mainly: \newline
1) a generalization of the use of the convolution: we make use of a set of filters
with several different sizes and we then combine all the convolved images in 
just one final significance image where objects of different scale lengths
are emphasized at the same time (see par \ref{sec:multifilt}). \newline
2) an optimization of the detection and photometry operations, that are 
performed at the same step using the final significance image (see par \ref{sec:candclass} 
and \ref{sec:artgal} for more details).

\subsection{Preliminary processing}

A very important issue when optimizing the detection of LSB
galaxies is ensuring that the sky is as flat as possible across
the image. Although the INT WFCS pipeline provides flat fielded
images, we use an automatic tool from the package Sextractor in
order to remove possible residual background fluctuations. The
technique consists of creating a map of the background sky by
interpolation of the mean values of pixels in a grid of the original image. The
grid size is chosen such as to preserve the biggest scale length
we want to detect with our convolution technique.
Noise reduction gained in this way is about $6\%$, but a flatter
background allows an easier and less contaminated application of
filters and a subsequent improved measure of source fluxes.

\subsection{Removal of stars and standard objects}

Before convolving the image with the set of filters, we remove all
the 'standard' astronomical objects, such as stars, very bright
galaxies, satellite tracks, hot pixels etc . This is necessary in
order to minimize the contamination of the sample by spurious objects, 
i.e. objects that could simulate LSB galaxies when convolved with the filters.

\begin{figure}
  \centerline{\epsfig{file=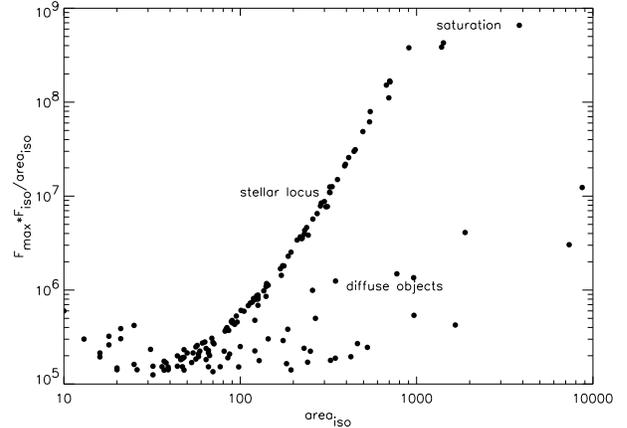,
height=6cm}}
  \caption{\footnotesize
{Sextractor detections for one of our typical images
 shown in the plane (isophotal area, surface isophotal flux
weighted by peak flux). It is clearly possible to discriminate the
stellar locus, a region of saturation and a region occupied by
diffuse objects such as galaxies}} {\label{fig:taglistelle}}
\end{figure}

The masking procedure is performed in two steps in which different
kinds of objects are considered: a first step is aimed at the
removal of big and bright objects, such as saturated stars or
bright galaxies, and a second one aimed at small stars. Although
Sextractor has the option of giving a cleaned output image, the
removal is not always effective and many stellar halos remain. We
have thus written our own code for bright or saturated objects and use
the Sextractor procedure for the few small stars left over after our
cleaning. Removal is performed by masking these objects with the
local median sky value with added Poissonian noise. We use
SExtractor just to detect all the standard objects in the image
and select the ones to reject using a criterium based on dimension
(isophotal area) and peak flux (surface isophotal flux weighted by
peak flux), that clearly discriminates a stellar locus, a possible
region of saturation and a region occupied by diffuse objects like
galaxies (fig. \ref{fig:taglistelle}).

Once the different regions in the plot are recognized, it is
possible to identify the objects we want to mask (fig.
\ref{fig:taglistelle}) \footnote{The theoretical trend of the
stellar locus is also easily representable assuming a gaussian
radial profile with a width equal to the psf}. The stellar locus
is fitted on each image so that the procedure is not affected by
any change in the seeing value. Once the stellar locus is fitted we 
consider a line with same slope, but constant coefficient lower of $1.5\sigma_{fit}$ 
and mask all the objects that are above this line. The object removal is performed
masking the region with the median sky value plus its Poissonian
noise. The size of the mask depends on the geometrical parameters
that Sextractor gives for each object (elliptical axis and Kron
radius), on peak intensity and on the radius at which the star
flux falls below $1\sigma_{sky}$.

Finally the removal of small stars left after this procedure, is
carried out using the star-subtracted image given by SExtractor.
We obtain, in this way, a 'star-cleaned' image.

The star-masking procedure decreases the useful area for galaxy
detection, so we take this into account when calculating numbers
per sq deg in the cluster.

\subsection{The multi-scale filter}\label{sec:multifilt}

\begin{figure}
  \centerline{\psfig{file=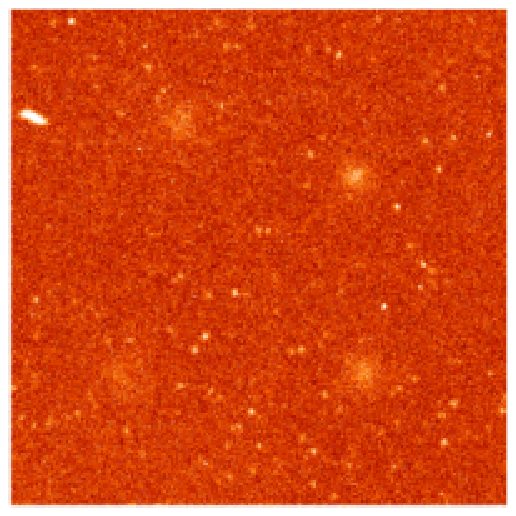,
height=6cm}}
\end{figure}
\begin{figure}
  \centerline{\psfig{file=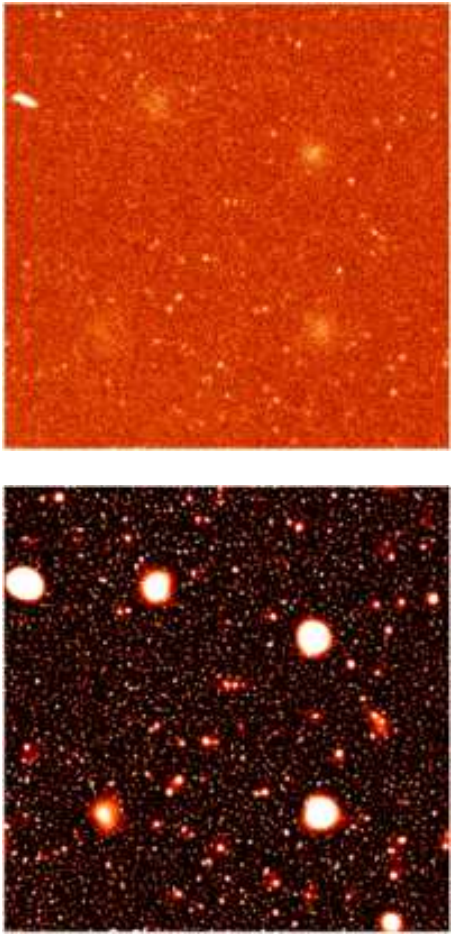,height=6cm}}
\caption{\footnotesize{ The top figure is a real image to which we
added 3 artificial galaxies of scale length (arcsec) and central
surface brightness (flux/sq arcsec) as follows:
(5,1$\sigma_{sky}$),(4,2$\sigma_{sky}$), (3,3$\sigma_{sky}$). The 2
top left detections are real galaxies in the image. \newline
Bottom figure is the final output of the algorithm it is clear
that the S/N ratio is improved and they are easily detectable.}}
{\label{fig:nsigmat}}
\end{figure}

The first problem when choosing the best filter for the detection
of LSB or dwarf galaxies is the determining of the optimal scale
size for the filter - the sizes of these galaxies are highly
variable. The necessity of selecting many different scales
requires the use of either a very wide band pass filter (with the
unavoidable consequence of including many different kind of
objects) or the application of many different filters where we
analyze the result from each one of them, requiring a huge working
time. Thus we decided to build a procedure in which we apply a
combination of filters of different sizes but we obtain (as
described below) just one final significance image. This final
image has the property of having each different size emphasized in
it at the same time. We then use it as a map of candidate
positions. 

The filters have exponential profiles and are
adjusted so that the convolution with a constant values, such
as an empty area of an image, gives zero as output. Each filter is
equal to zero for $r=3\alpha$ and $r=6\alpha$ (which means that it
weights as positive everything inside $r=3\alpha$ (typical size of
an exponential object of scale length $3\alpha$) and subtracts
whatever is between $r=3\alpha$ and $r=6\alpha$). In this way we
obtain a scale selective filter: everything that is smaller or
bigger than the filter scale size is severely dimmed.

The cleaned image is convolved with each filter and the output is
a series of convolved images. Each of them improves the signal to noise
ratio (S/N) of objects with scale sizes matching the filter scale. Using these
convolved images we build a final output image whose value in each
pixel is equal to the maximum value assumed in the stack of
convolved images \footnote{ The intensity on each convolved image
is measured as multiples of the convolved image noise, so that the
measure of intensities on different convolved images can be
compared}. The main property of this final image is that objects
corresponding to all the different sizes of filters are emphasized
at the same time (fig \ref{fig:nsigmat}).

The final image is used as a map for the positions of candidate
galaxies; both detection and photometry are then performed on it.
It is a {\it significance image} (as the value of each pixel is
expressed in multiples of noise of the convolved image
correspondent to the best matching filter) and the detection of
the candidates is done taking all peaks above a $3\sigma$
threshold (see sec. \ref{sec:candclass}). On a different array we
save the filter scale on which each maximum for each pixel is
found. That filter scale corresponds to the best matching scale,
and this is our estimate of the scale length of the object.


From a computational point of view, once the filter scale sizes
are decided, the set of filter arrays are built just once and then
restored in the code at the convolution step. As a first
application, we use filters with a radial symmetry, but the code also
allows for the possibility of using an elliptical symmetry, which
is optimized for edge on or intrinsically elongated objects.

The peak value on the output image also contains all the information we need for the
photometry of the object, as shown in the following section.

\subsection{Candidate classification: parameter estimation}\label{sec:candclass}

Given the properties of the filters (see section \ref{sec:multifilt}) it can be shown that
when the scale size of the filter $H$ matches the scale size of the galaxy $h$ 
(which is the filter scale for
which the value of the convolution is found to be maximum), the value of
the convolution integral is:
\begin{equation}
\centerline {$ I=\int f_{0} e^{-\frac{r}{h}}(F_{*}e^{-\frac{r}{H}})2\pi
rdr= \frac{F_*}{2}\pi f_0 h ^2 $}
\end{equation}
\noindent where $f_{0} e^{-\frac{r}{h}}$ is the assumed
exponential profile of the galaxy and $F_{*}e^{-\frac{r}{H}}$
is just the positive part of the filter, as the object has zero
flux on the negative part. Knowing the best matching scale, which
corresponds to our best estimation of the object scale length, we
can estimate the central original flux as:
\begin{equation}
\centerline{ $ f_{0}=\frac{I}{2\pi h^{2}_{\small best}F_{*} }$}
\label{eqn:fzero}
\end{equation}

\noindent and calculate the central surface brightness.

In the ideal case of a poissonian distribution for pixels in the
image and assuming pixels to be uncorrelated, we can derive a
simple relationship between the noise of the original image
($\sigma_{i}$) and the noise in each convolved image
($\sigma_{o}$). Within these assumptions, for a convolution with a
simple $n\times n$ box filter, the noise reduction is:
\begin{equation}
\centerline {$ \sigma_{o}=\frac{\sigma_{i}}{\sqrt{n}}$}
\end{equation}
\noindent
The relation still holds for the exponential filters and
in general we can then write:
\begin{equation}
\centerline {$ \sigma_{o}=k(H)\sigma_{i}$} \label{eqn:noise}
\end{equation}
where the coefficient k is different for each filter and is
related to the filter scale size H. \newline Using eqs.
\ref{eqn:fzero} and \ref{eqn:noise} we can then show how the
detection threshold on the final convolved image
($t=N_{t}\sigma_{o}$) relates to a threshold in the minimum
central surface brightness which is detected for each scale
length:
\begin{equation}
\centerline {$ f_{0,min}=\frac{t}{2\pi h^{2}F_{*}}=\frac{N_{t}k(h)}{2\pi h^{2}F_{*}}\sigma{i}$}
\label{eqn:f0noise}
\end{equation}

However the assumption that pixels are uncorrelated is not
completely true (because of PSF) and these relationships need to
be calibrated for our data. In order to calibrate them we used
artificial galaxies added to real images. The results of these
simulations are shown in the following section.

\subsection{Artificial galaxy simulations}\label{sec:artgal}

In order to test the efficiency of the method, we ran simulations
using artificial galaxies added to real images. The use of
artificial galaxies allows us to test the algorithm, exploring
systematically the parameter space of scale length (h) and
central surface brightness ($\mu_{0}$) and thus total magnitude.
The added artificial galaxies have exponential profiles (as
observations indicate that this is the best representation for
dwarf and LSB galaxies profiles. Davies et al., 1988), are
convolved with a gaussian that simulates the seeing and are added
to the real data with their poissonian noise. The use of these
simulations allows us to determine the efficiency for detection
and photometry of the objects.

\begin{figure}
  \centerline{\epsfig{file=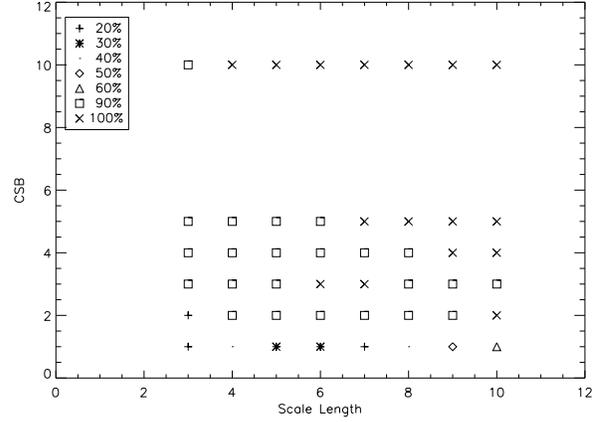,height=6cm}}
  \caption{\footnotesize{
    Detection efficiency plotted as a function
    of scale length $h$ and $\mu_{0}$. The CSB ($\mu_{0}$) is measured as a
    multiple of the sky noise and in our images this correspond to: $1\sigma\sim26.1$mag/sq arcsec,
    $2\sigma\sim25.3$mag/sq arcsec,$3\sigma\sim24.9$mag/sq arcsec,$4\sigma\sim24.6$mag/sq arcsec,
    $5\sigma\sim24.3$mag/sq arcsec,$10\sigma\sim23.6$mag/sq arcsec. As shown in the plot,
    different symbols refer to different efficiencies.}}
{\label{fig:efficiency}}
\end{figure}

In fig. \ref{fig:efficiency} we plot the efficiency of detection
as a function of scale length and central surface brightness of
the artificial galaxies, where different symbols refer to
different efficiencies. The efficiency is very high ($90-100\%$)
over almost all of the simulated region. As expected the
efficiency drops for small and faint objects because of their low S/N. 
We now have an estimate of completeness and contamination of our detection method
and we can correct for these effects. The objects that we can 
detect are just the objects we want to look for in Virgo. The
faintest and smallest galaxies we are able to detect correspond to
$M_{B} \sim -9.5$ and the  brightest and biggest to $M_{B} \sim
-14.5$.

\begin{figure}
  \centerline{\epsfig{file=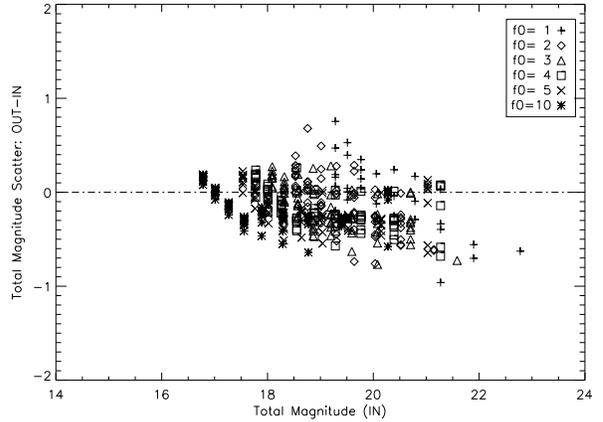,
height=6cm}}
  \caption{\footnotesize{
    The measured magnitude minus input one is plotted against input total
magnitude.
    Different symbols refer to different values of the $\mu_{0}$. The biggest errors
    are for very faint objects.}}
{\label{fig:mutotscatt}}
\end{figure}

Once detected, and knowing the best matching filter scale, we can
determine the central surface brightness (using eq
\ref{eqn:fzero}) and thus the total magnitude of the objects. In
figure \ref{fig:mutotscatt} we show the difference between input
values and recovered ones for the total magnitudes of the
simulated galaxies. The efficiency in recovering the object's
scale length is strongly dependent on how big the gap is between
the scale size of different filters. In order to obtain a good
sampling of the scale lengths that we expect for dwarfs in Virgo,
we decided to use the following 2,3,4,5,6,7 and 9 arcsec filter
scales. Although we used smaller filters the final minimum scale
size for objects in our Virgo sample is 4 arc sec (see sec.
\ref{sec:numsim} for comments on this).

Although we do have some scatter in the recovered scale length and
central surface brightnesses, the two compensate to give an
estimation of the total magnitude with mean error of $\pm 0.5$.

The standard way of measuring photometric parameters is by fitting to the radial surface brightness profile.
 The detection and fitting are two distinct operations. Using our method we maximize our detection
 efficiency by detecting the entire image, rather than just its poor S/N edge, and we also obtain the
best fitting parameters at the same time.


\section{First results: The Luminosity Function }

\subsection{The radial number density profile of the cluster} \label{sec:numbdens}

Clusters are very interesting regions to study in order to
understand the role played by the environment on galaxy formation
and evolution. They are also important because it might be
expected that the cluster population surface density decreases
with radius from the cluster centre and eventually merges into the
field. This would imply that the properties of galaxies in the
outskirts of the cluster must be linked to the ones of the
field population.

The sample of galaxies that we have detected with our technique
extends from the centre of the cluster (identified as M87) outward
for 7 degrees. Over the resulting area of $\sim 14$ sq degrees we
have identified 105 new extended dwarf LSB previously uncatalogued
(see VCC, Impey et al. 1988, Trentham et al. 2002). 

\begin{figure}
  \centerline{\epsfig{file=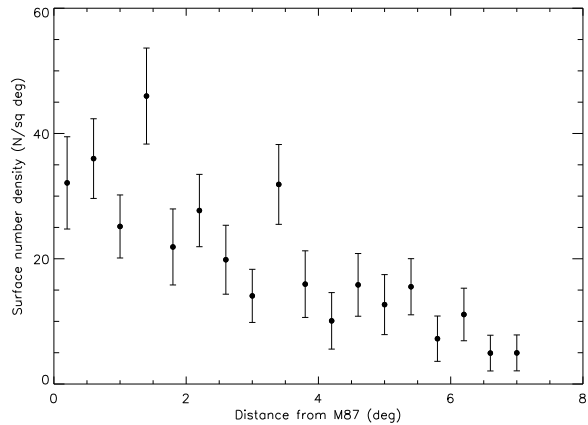,
height=6cm}}
  \caption{\footnotesize{
In this figure we plot the number density profile of the galaxies
detected within our selection criteria (h$\geq 4$,
$\mu_{0}\geq 23$) versus distance from the cluster centre
(identified as M87). As expected for a cluster member population
the density decrease when going further from the centre and
eventually drops to almost zero approaching the cluster edge.}}
{\label{fig:numbdens}}
\end{figure}

Before discussing the
implications on the luminosity function, though, we need to
demonstrate that we have a Virgo cluster sample rather than a
sample contaminated by background (or foreground) objects. \newline
Background contamination has been one of our main concerns. In
fig. \ref{fig:numbdens} we have plotted the surface number density
of our detections against cluster radius. This is the raw data:
there are no corrections for contamination or completeness in this
plot. As expected for a cluster member population the density
decreases when going further from the centre and eventually drops
to an almost constant value close to zero, approaching the cluster edge. 
If our sample were highly contaminated by background galaxies, we would expect an
almost flat distribution of galaxies, not one dependent on
distance from the centre of the cluster.

An exponential plus a constant (background) fit to the
distribution gives a scale length of $2.2^{\circ} \pm 0.2^{\circ}$ (at the
distance of the Virgo Cluster this corresponds to 0.7 Mpc) 
and a background galaxy density of $5 \pm 1$ gal per sq deg. 
This is consistent within the errors with the value of 4 gal/sq deg 
found for the background in the offset fields
described in section 3.2 and can thus be considered as the
non-members contamination in our sample. In summary, although there might 
still be some contamination by background galaxies, we demonstrated it is minimal
(see our numerical simulations, the offset fields number counts and 
fig \ref{fig:numbdens}) 

\begin{figure}
  \centerline{\epsfig{file=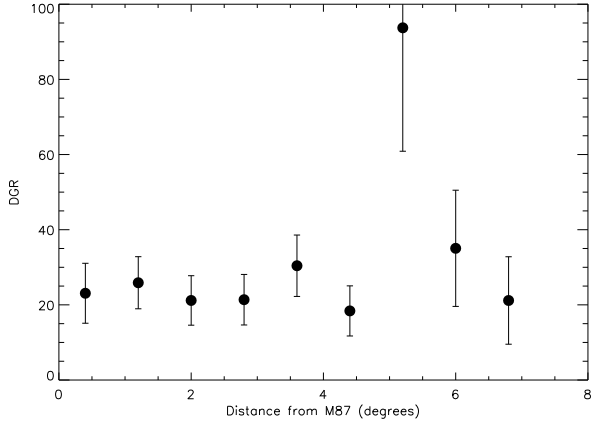,
height=6cm}}
  \caption{\footnotesize{
Dwarf to giant ratio as a function of distance from M87. Here we
define dwarf the galaxies with $-14 \leq M_{B} \leq -10$ and giant
the ones with $M_{B} \leq -19$. The peak at $\sim 5^{o}$ is mainly
due to a dip in the distribution of giant and not to an excess of
dwarfs.}} {\label{fig:dgr}}
\end{figure}

This radial distribution can be compared with that of the bright
galaxies. We have defined a Dwarf-to-Giant Ratio (DGR) which we
will use in subsequent papers for comparison with different
environments. This is the ratio of dwarf galaxies, definided as
those with $-14 \leq M_{B} \leq -10$, to giant galaxies with
$M_{B} \leq -19$. We use this quantity because in some
environments there are too few galaxies to construct a luminosity
function. \newline In fig \ref{fig:dgr} we show this ratio as a
function of distance from M87. Interestingly this ratio remains
rather flat with a median value of $\sim 20$: the dwarfs number and
giants one decline with clustercentric distance with the same
scale length, resulting in a constant DGR.
\newline Sabatini S., Roberts S. and Davies J. (JENAM, 2002) have
shown evidence that the DGR is about 4 for the field population.
This is just about the number you would obtain with our
selection criteria if observing the Milky Way from the distance of 
the Virgo cluster. Out of
the group of galaxies that Mateo (1998) assigns to the Milky Way
just the dwarfs Sextans Fornax and Sagittarius would meet our
selection criteria. Again this would give a value for the DGR of
3, lower than in Virgo. As most dwarf galaxies in clusters are probably 
not bound to individual giant galaxies we can also do the same test for all the dwarfs
in the Local Group, in order to compare it with the Virgo Cluster. Again we obtain 
a DGR of 4.
\newline Also, interestingly, the
Virgo cluster DGR does not appear to smoothly blend into the
field; if it did we would expect DGR to gradually decrease to the
value in the field. Note also that we must be close to the 'edge'
of the cluster because the galaxy counts are about the same as
those in our offset fields. Thus the Virgo cluster
environment seems to be very different to that of the field even
in its most outer regions.

\subsection{The faint end slope of the Luminosity Function.}
In this paper we are primarily interested in the number of dwarf
galaxies in the cluster. When sufficient galaxies are available
this is found by fitting to the faint end of the luminosity
function. Our method enables us to detect galaxies with the
following range of absolute magnitudes: $M_{B}$=-10
(h$_{min}=4''$, $\mu_{0}^{min}=26$ $B\mu$) to $M_{B}=-15$
($\mu_{0}^{max}=23$, h$_{max}=9$).

\begin{figure}
  \centerline{\epsfig{file=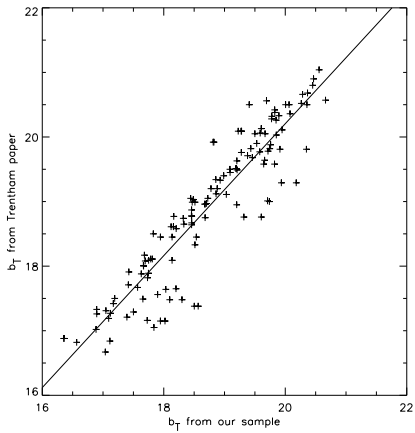,
height=6cm}}
  \caption{\footnotesize{
Comparison between apparent magnitudes as measured with our method (x axis) and 
those in Trentham et al. 2002. A linear fit to the data gives a slope of 1.02
and a constant of -0.22.
}} {\label{fig:magcomp}}
\end{figure}

An important check is to compare our derived magnitudes with galaxies 
common to previous catalogues of the Virgo Cluster. We have 143 
galaxies in our sample that are listed in Trentham et al. 2002 and we show a plot of 
our measured apparent magnitudes against the Trentham et al. measurements 
in fig \ref{fig:magcomp}: a 
linear fit to this plot results in a slope of 1.02 and a constant value of -0.22. 
Our magnitudes tend to be slightly brighter then Trentham's, but the difference
lies within the errors we expect in our measurements (see fig \ref{fig:mutotscatt}). 
Galaxies listed in Trentham et al. 2002 that are not in our 
sample have been checked and they are either too bright to fall within our magnitude range 
or they lie in masked regions of the images (in our calculations
we take into account the area lost due to the removal of stars). 

\begin{figure}
  \centerline{\epsfig{file=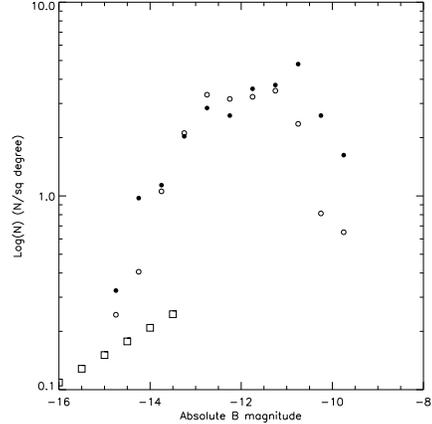,
height=6cm}}
  \caption{\footnotesize{
In this figure we plot the faint end of the luminosity function.
Open circles are counts for the raw data, while filled points are
corrected for detection efficiency and background contamination.
Squares are points from the VCC luminosity function with slope
$\alpha = -1.35$, renormalized for our survey area.
 }}
{\label{fig:LF}}
\end{figure}

Our derived luminosity function is shown in fig. \ref{fig:LF}. We show the raw 
data and the corrected data. The number of objects detected in each bin 
of $h$ and $\mu_{0}$ has been corrected for:
\begin{enumerate}
\item background
contamination (as obtained from our numerical simulations of a cone of the universe).
\item incompleteness
(as estimated by the detection efficiency of the algorithm with artificial galaxies).
\end{enumerate}
The corrections make little difference to the numbers detected in each bin.
Assuming that the drop-off in numbers beyond $M_{B}=-10.5$ is due to incompleteness
 we have fitted the luminosity function in the
magnitude range -14.5 to -10.5. This gives a value for
the faint end slope of $-1.7 \pm 0.2$ for the raw counts and $-1.6 \pm 0.1$ for the counts
corrected for incompleteness.

\begin{figure}
  \centerline{\epsfig{file=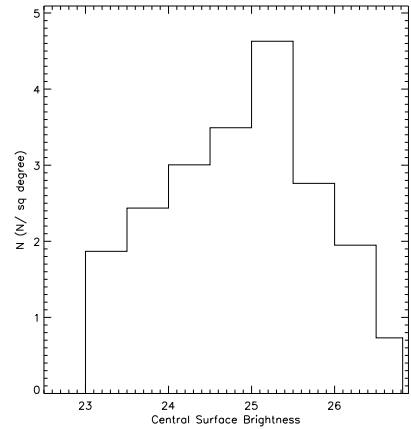,
height=6cm}}
  \caption{\footnotesize{ Histogram of Central Surface Brightness distribution for
our detections.
 }}
{\label{fig:sbdistrib}}
\end{figure}

It is difficult to combine galaxy counts from samples selected in different ways. For 
example in Kambas et al. (2000) we show how the selection of three different samples of 
galaxies leads to a disjoint surface brightness distribution because in each case the 
selection criteria preferentially picks galaxies of a given surface brightness. The 
surface brightness distribution of this sample (fig.
\ref{fig:sbdistrib})
is also peaked.

Even so in fig. \ref{fig:LF} we have also shown, for comparison, the number counts for
the VCC \footnote{These can also be used to compare our data with the number counts obtained in
Trentham et al 2002, as his findings fit very well with the Luminosity Function of the VCC
(see their fig 5).}. Including the VCC data and fitting the faint
end slope between -16 and -10.5 we obtain a value of -1.8 for
$\alpha$ \footnote{We can't write an error on this value because we don't have a 
completeness function for the VCC catalogue.}. 
The values that we obtain are in agreement with a
steepening of the LF when including in the sample the contribution
from faint and LSB galaxies. The original VCC value of -1.35
(Sandage et al., 1984) had already been brought into question
by Binggeli et al. (1988) who suggested a steepening to -1.7
when including a possible dwarf population that had escaped
detection. The final slope we obtain is not as steep as the one
obtained by Phillipps et al. (1998) using the background
field subtraction method. However, in order to compare our results 
with Phillipps et al., we should consider our raw data counts 
(as they didn't make any correction for incompleteness) that result
in a slope of $-1.7 \pm 0.2$. Neglecting the last point in their LF, that
might be highly background contaminated (S. Phillipps, private communication),
 their slope is $\sim -1.9 \pm 0.15$. This then is consistent with our result.

\begin{figure}
  \centerline{\epsfig{file=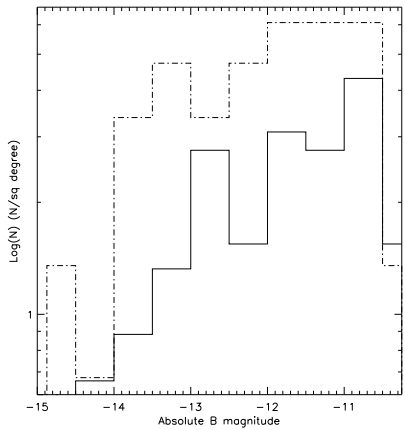,
height=6cm}}
  \caption{\footnotesize{
In this figure we plot the faint end part of the
luminosity function separating
the cluster into an inner (distance from M87 $\leq 0.8^{\circ}$) and outer(distance from M87 
$\geq 1.6^{\circ}$)
region. Dashed-dotted line refers to the former, while filled line to
the latter. Although the
numbers give a poor statistics, a K-S test shows that the two
distributions are different with a
$90\%$ probability.
A fit to the same range of magnitudes as for the total LF, a steeper
value for the faint end slope
of the LF in the outer region compared with the inner one ($-1.8 \pm 0.2$ and $ -1.4 \pm 0.2$ )
 }}
{\label{fig:inoutlf}}
\end{figure}

We can also calculate a separate luminosity function for the inner
and outer region of the cluster (see fig. \ref{fig:inoutlf}).
Dashed-dotted line refers to the former, while the filled line to
the latter. We have chosen these regions because they correspond
roughly to that part of the cluster that is within the virial
radius and that which is outside it. Galaxies within the virial
radius should have been much more affected by interactions with
other cluster galaxies than those in the dynamically unrelaxed
outer regions (Bohringer, 1995). A fit to the same range of
magnitudes as for the total luminosity function results in a
steeper value for the faint end slope in the outer region compared
with the inner ($-1.8 \pm 0.2$ and $-1.4 \pm 0.2$ ). A K-S test
shows that the two distributions are different with a $90\%$
probability. If confirmed by a larger statistics, this result is
consistent with the idea that the faintest galaxies are more
abundant in the outer regions of clusters, while in the denser
inner regions they have partly been accreted by larger galaxies or
have dimmed or even been disrupted by tidal interactions.

\subsection{Total Luminosity and Mass}

The total light (corrected for the area of the cluster sampled
compared to the VCC) due to the population of dwarf LSB galaxies
of our sample is

\begin{equation}
L_{Dwarf}= 4 \times 10^{9} L_{\odot}
\end{equation}

which is just $1/50$ of that due to VCC galaxies. So this
population of galaxies contributes only a small fraction of the
light contributed by the bright galaxies.

A surface brightness level of about 28 B mag/sq arcsec has
recently been predicted for intra-cluster light from stars
associated with intra-cluster planetary nebulae (Arnaboldi et al.
2002). The average integrated surface brightness from the galaxies we detected
is $\mu_{tot} \approx 32$ mag/sq arcsec in the inner region of
the cluster (within $0.8^{o}$ from M87). This is far too faint to
have been previously detectable as a surface brightness
enhancement. This, combined with the planetary nebulae data, may
indicate that there are more even lower surface brightness
structures to be discovered between the galaxies.

Despite the small contribution to the light, dwarf galaxies may
contribute a larger fraction to the mass. Recently very large
mass-to-light ratios have been found for Local Group galaxies (see
for example Mateo 1998 and particularly Kleyna et al. 2002).
Assuming a mass-to-light relation as in Davies
et al. (2002):

\begin{equation}
\frac{M}{L}= 10^{3.5}L^{-0.25}
\end{equation}

\noindent the total mass due to the dwarf LSB population over an area of $\sim 14$ sq deg is

\begin{equation}
M_{Dwarf}= 1.9 \times 10^{11} M_{\odot}
\end{equation}

Rescaled to the total area of the cluster, this is $1/10$ of the total mass from the 
VCC galaxies. Given that other
low surface brightness material associated with intra-cluster
stars exists, it is possible that there is as much mass in the low surface
brightness component of the Virgo cluster as there is in the
brighter galaxies.

\section{Conclusions}

Observations of the relative numbers of dwarf galaxies in
different environments present a strong challenge to galaxy
formation models that predict large numbers of dwarf galaxies in
all environments. They also present a challenge to those models
that predict global suppression of dwarf galaxy formation. The
Virgo cluster is a very different environment from that of the
Local Group and from that of the general field. It has a very
large number of dwarf galaxies compared to the giant galaxy
population.

In this paper we have described a new automated technique for
finding low surface brightness galaxies on wide field CCD data. We
have carried out simulations to ensure that our detection method
and selection criteria enable us to preferentially select cluster
galaxies, rather than those in the background. From the decrease
in surface number density with clustercentric distance we believe
that we have achieved this goal. 

Over the resulting area of $\sim 14$ sq degrees that we analysed, we
have identified 105 new extended dwarf LSB previously uncatalogued
(see VCC, Impey et al. 1988, Trentham et al. 2002). 
The resulting luminosity function is considerably steeper than that 
inferred from an extrapolation of the data in the VCC. 
The cluster luminosity function appears to be
steeper in the outer parts of the cluster than in the inner part,
though the dwarf-to-giant ratio remains almost constant. Although
these galaxies contribute only a small fraction of the luminosity
of the cluster they may contribute significantly to the galactic
mass of the cluster, given recently measured large mass-to-light
for Local Group dSph galaxies.

\bsp

\label{lastpage}


\begin{thebibliography}{99}

\bibitem{b1} Armandroff T.E., Davies J.E., Jacoby G.H., 1998, AJ, 116, 2287
\bibitem{b1} Arnaboldi M., Aguerri A.L., Napolitano N., Gerhard O., Freeman K., et al., 2002, AJ, 123, 760A
\bibitem{b2} Bernstein G.M., Nichol R.C., Tyson J.A., Ulmer M.P., Wittman D., 1995, AJ, 110, 1570B
\bibitem{b3} Bertin E., Arnouts S., 1996, A\&AS, 117, 393B
\bibitem{b4} Binggeli B., Sandage A., Tarenghi M., 1984, AJ, 89, 64
\bibitem{b5} Binggeli B., Tammann G.A., Sandage A., 1985, AJ, 94, 251
\bibitem{b6} Binggeli B., Sandage A., Tammann G.A., 1988, ARAA, 26, 509
\bibitem{b7} Binggeli B., Sandage A., Tammann G.A., 1985, AJ, 90, 1681B
\bibitem{b8} Blanton M.R., Dalcanton J., Eisenstein D. et al., 2001, AJ, 121, 2538
\bibitem{b9} Bohringer H., 1995, Annals of the New York Accademy of Science, 759, 67
\bibitem{b9} Bothun G.D., Impey C.D., Malin D.F., 1991, ApJ, 376, 404 
\bibitem{b10} Bullock J.S., Kravtsov A.V., Weinberg D.H., 2000, ApJ, 539, 517
\bibitem{b11} Conselice C.J., Gallagher J.S., Wyse R.F.G., 2001, ApJ, 559, 791
\bibitem{b12} Davies J., Phillipps S., Cawson M., Disney M., Kibblewhite E., 1988, MNRAS, 232, 239D
\bibitem{b13} Davies J., Linder S., Roberts S., Sabatini S., Smith R. and Evans R., 2002, MNRAS, submitted
\bibitem{b14} Dekel A., Silk J., 1986, ApJ, 303, 39D
\bibitem{b15} Driver S., 1999, AJ, 526L, 69D
\bibitem{b16} Efstathiou G., 1992, MNRAS, 256P, 43E
\bibitem{b16} Flint K., Metevier A.J., Bolte M., De Oliveira C.M., 2001, ApJSS, 134,53
\bibitem{b17} Fouque P., Solanes J.M, Sanchis T., Balkowski C, 2001, A\&A, 375, 770
\bibitem{b18} Gavazzi G., Bonfanti C., Sanvito G. et al., astro-ph/0205074, ApJ, in press
\bibitem{b19} Impey C., Bothun G., Malin D., 1988, ApJ, 330, 634
\bibitem{b20} Kambas A., Davies J., Smith R., Bianchi S., Haynes J., 2000, AJ, 120, 1316
\bibitem{b21} Kleyna J., Wilkinson M.I., Evans N.W., Gilmore G., Frayn C., 2002, MNRAS, 330, 792K
\bibitem{b22} Lahav O. \& the 2dFGRS team, 2002, the 5th RESCEU Symposium, Tokyo, Universal Academy Press
\bibitem{b23} Liske J., Lemon D.J., Driver S.P. et al., astro-ph/0207555
\bibitem{b24} Madgwick D., Lahav O. et al., 2001, MNRAS, 333, 133M
\bibitem{b25} Mateo M.L., 1998, ARA\&A, 36, 435M
\bibitem{b26} Moore B., Lake G., Quinn T., Stadel J., 1999, MNRAS, 304, 465M
\bibitem{b27} Morgan I., Smith R.M., Phillipps S., 1998, MNRAS,295, 99M
\bibitem{b28} Morshidi-Esslinger Z., Davies J., Smith R., 1999, MNRAS, 304
\bibitem{b29} Phillipps S., Parker Q., Scwhartzenberg J., Jones J.B., 1998, ApJ 493, 59L
\bibitem{b30} Phillipps S., Driver S.P., Couch W.J., Smith R.M., 1998a, ApJ, 498, L119
\bibitem{b31} Russell J.S., Lucey J.R., Hudson M.J., Schlegel D.J., Davies R.L., 2000, MNRAS, 313, 469S
\bibitem{b32} Sabatini S., Scaramella R., Testa V., Andreon S., Longo G., Djorgovsky G.,
        De Carvalho R.R., 1999, SAIt proceedings
\bibitem{b33} Sabatini S., Roberts S., Davies J., 2002, JENAM proceedings
\bibitem{b34} Sandage A., Binggeli B, 1984, AJ, 89, 919S
\bibitem{b35} Sandage A., Binggeli B., Tammann G., 1985, AJ, 90, 395S
\bibitem{b36} Sandage A., Binggeli B., Tammann G., 1985a, AJ, 90, 1759S
\bibitem{b37} Schombert J.M., Bothun G.D., 1988, AJ, 95, 1389S
\bibitem{b38} Shibata R., Matsushita K., Yamasaki N.Y. et al., 2001, ApJ, 549, 228S
\bibitem{b39} Shapley H., Ames A., 1932, Annals of Harvard College Observatory, Cambridge, Mass
\bibitem{b40} Smith R.J., Lucey J.R., Hudson M.J., Schlegel D.J., Roger L.D., 2000, MNRAS, 313, 469S
\bibitem{b41} Tikhonov N.A., Galazutdinova O.A., Drozdovskii I.O., 2000, Ap, 43, I 4
\bibitem{b42} Trentham N., Tully B., Vereijen M., 2001,  MNRAS , 325, 385
\bibitem{b43} Trentham N., Hodgkin S., 2002, MNRAS, 333, 423T
\bibitem{b44} Trentham N., Tully R.B., 2002, MNRAS, 335, 712T
\bibitem{b45} Tuffs R., Popescu C., Pierini D. et al., 2002 ApJS, 139, 37T
\bibitem{b46} Tully R.B., Somerville R.S., Trentham N. et al, 2002, ApJ, 569
\bibitem{b47} Valotto C.A., Moore B., Lambas D.G., 2001, ApJ, 546, 157V
\bibitem{b48} Van Driel W., Ragaigne D., Boselli A. et al., 2000, A\&AS, 144, 463V

\end{thebibliography}
\end{document}